\documentclass[prb, twocolumn, showpacs, showkeys] {revtex4}
\usepackage{latexsym}
\usepackage{graphics}
\usepackage{dcolumn}

\begin{document}

\title{Phonon anomaly at the charge ordering transition in
\textit{1T}-TaS$_2$}

\date{\today}

\author{L.V. Gasparov}
\author{K.G. Brown}
\affiliation{University of North Florida, Department of Chemistry and Physics,
4567 St.Johns Bluff Rd. South\\
Jacksonville Fl 32224}
\author{A.C. Wint}
\author{D.B. Tanner}
\affiliation{University of Florida, Department of Physics, PO Box
118440, Gainesville, Fl 32611-8440}
\author{H. Berger}
\author{G. Margaritondo}
\author{R. Ga\'{a}l}
\author{L. Forr\'{o}}
\affiliation{Institute of Physics of Complex Matter,
\'{E}cole Polytechnique F\'{e}d\'{e}ral de Lausanne,\\
CH-1015 Lausanne, Switzerland}

\begin{abstract}
The infrared reflectance of the transition metal chalcogenide
\textit{1T}-TaS$_2$ has been measured at temperatures from 30~K to
360~K over 30-45,000~cm$^{-1}$ (4~meV-5.5~eV). The optical
conductivity was obtained by Kramers-Kronig analysis. At 360K only
modest traces of the phonon lines are noticeable. The phonon modes
are followed by a pseudogap-like increase of the optical
conductivity, with direct optical transitions observed at
frequencies above 1eV. As the temperature decreases,  the low
frequency conductivity also decreases, phonon modes become more
pronounced and pseudogap develops into a gap at 800~cm$^{-1}$
(100~meV). We observe an anomalous frequency dependence of the
208~cm$^{-1}$ infrared-active phonon mode. This mode demonstrates
softening as the temperature decreases below the 180~K
metal-to-insulator transition. The same mode demonstrates strong
hysteresis of the frequency and linewidth changes, similar in its
temperature behavior to the hysteresis in the dc-resistivity. We
discuss a possible relation of the observed softening of the mode
to the structural changes associated with the metal-to-insulator
transition.

\end{abstract}


\maketitle

\section{Introduction}

\subsection{Transition metal chalcogenides}

Transition metal chalcogenides (TMC) form a  group of compounds
which can be characterized by an X-M-X structure, where X is a
chalcogen and M is a transition metal.\cite{Wilson69} The majority
of TMC are layered compounds with a hexagonally-packed metal layer
sandwiched between two chalcogen layers making an X-M-X sandwich.
These three-layer sandwiches are weakly bound to each other by
weak force, leading to a quasi-two-dimensional behavior in these
compounds.

\begin{figure}
\includegraphics{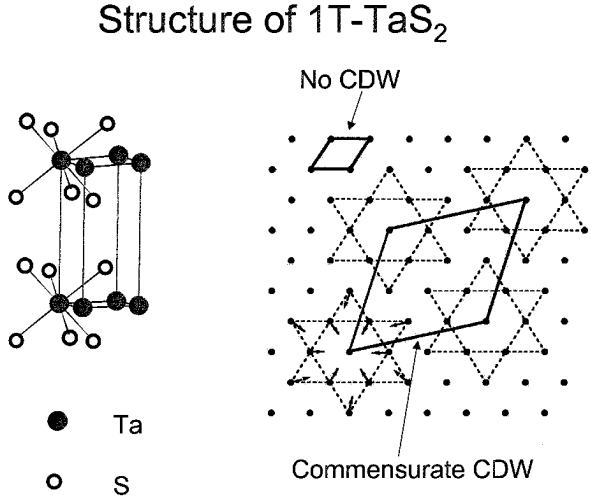}
\caption{\label{fig1} Structure of
\textit{1T}-TaS$_2$. Left-hand panel shows the unit cell in the
incommensurate state. The upper part of the right-hand panel shows
the arrangement of Ta atoms when no CDW is present. The bottom
part of the right-hand panel demonstrates ``star of David" cluster
formation at 180~K transition. Small arrows indicate the
displacement of Ta atoms when this cluster is formed. }
\end{figure}

Different  arrangements between layers within the sandwich lead to
a variety of so-called polytypic modifications. These are
indicated by the corresponding prefix to the name of compound. In
particular, the pure octahedral coordination of the metal atom by
chalcogen atoms is denoted by \textit{1T}, the pure trigonal
prismatic coordinations are denoted by \textit{2H, 3R} or
\textit{4Hc}, and  mixed coordinations are denoted by \textit{4Hb}
or \textit{6R} \cite{Wilson69}. Quasi-two-dimensional structure of
these materials results in a variety of unique properties. In
particular \textit{1T}-TaS$_2$, which is the subject of this
report, was one of the first compounds where a charge density wave
(CDW) was experimentally observed.\cite{Wilson69} In this compound
the tantalum atoms are octahedrally coordinated by sulphur atoms
(see Fig.\ref{fig1}, left panel).

The development of the CDW in \textit{1T}-TaS$_2$ yields two
metal-to- insulator (MI) transitions at 350~K and 180~K, with the
latter transition demonstrating hysteresis. When the crystal is
cooled down the transition is at $\approx$ 180~K, whereas when it
is warmed up from temperatures below 180K the transition occurs at
$\approx$ 230K.

The physics of this compound can be understood in terms of the
interaction between  the CDW and the crystal lattice. MI
transitions separate the phase diagram into three regions. At
temperatures above 350K there is an incommensurate  charge density
wave, which becomes nearly commensurate at 350~K. This process is
accompanied by the first MI transition and the formation of a
domain structure. Finally at the second MI transition the CDW
becomes completely commensurate. The hysteresis of the second
transition is related to the domain structure.

\textit{1T}-TaS$_2$ has been intensively studied with many
different methods. Recently, the interest in this material was
revived mainly due to substantial improvements in angle-resolved
photoemission spectroscopy (ARPES). In their ARPES studies of this
compound Pillo et al.\cite{Pillo01} observed a rich electronic
structure near the Fermi level, with a strong indication of the
opening of the gap of $\approx$ 180~meV below the 180~K MI
transition.

There is a variety of early infrared data on this material.
\cite{Wilson69,Barker75,Lucovsky76,Karecki76,Karecki79,Uch81}
Since then, progress in infrared spectroscopy equipment has
allowed for better sensitivity and resolution, as well as for
broader frequency range measurements.


None of the existing  publications gives a detailed analysis of
the  temperature dependance of the frequencies and linewidths of
the infrared-active phonons. Neither is there  a discussion of the
effect of the hysteresis of the 180~K transition on optical data.

We undertook a  systematic study of the \textit{1T}-TaS$_2$
optical reflectance in the frequency range from far infrared to
ultraviolet for the temperature range from 30 to 360~K. The effect
of the hysteresis on the optical properties of the compound has
been also addressed. In this short report we discuss our findings
in the low-frequency range (below 1000~cm$^{-1}$) with the main
emphasis on the changes in phonon spectrum of the compound.

\subsection{The charge density wave and
 optical phonons in \textit{1T}-TaS$_2$}

Above the 350K-transition, the  crystal structure of
\textit{1T}-Ta$S_2$ can be described as a CdI$_2$-type structure
belonging to the space group  $D_{3d}^d,~P~\bar3~2/m~1$. The unit
cell contains one formula unit (3 atoms). As a result, one should
expect 9 zone-center phonon modes. Group theory predicts the
following symmetry in the modes.\cite{Uch81}

\[A_{1g}+E_{g}+2A_{2u}+2E_{u}\]
There are two Raman-active modes ($A_{1g}+E_{g}$) and two
infrared-active modes ($A_{2u}+E_{u}$). The two remaining modes
are acoustic phonons. Note that E-type modes are double
degenerate.

Below 350~K the CDW becomes nearly commensurate, leading to
domains with different orientations of the CDW. Finally at 180~K
when the crystal is cooled down, the  CDW becomes completely
commensurate. This process is accompanied with the formation of
the so-called
$\sqrt{13}$\textit{a}$\times\sqrt{13}$\textit{a}$\times$13\textit{c}
``star of David" cluster. The right panel of Fig.\ref{fig1}
illustrates this process. In particular, the Ta atoms in the
center  of the ``star" become the corners of a new unit cell. The
nearest neighboring Ta atoms move toward the ``star"-center Ta
atoms, and next nearest neighboring  Ta atoms also move toward it.
The displacement of the Ta atoms is accompanied by  bulging of the
S layers\cite{Wilson75,Fazekas79}. The new unit cell has a
$C_{i}^{1},~P\bar1$  symmetry. It contains 39 atoms and thus leads
to a total of 117 modes. Polarized Raman measurements of Uchida
et.~al\cite{Uch81} displayed the selection rules  between A$_{1g}$
and E$_{g}$ modes. It was argued that the symmetry of a single
layer ($C_{3i}^{1},P3$) can better describe this behavior.
Following Uchida\cite{Uch81} we expect 117 normal modes.

\[19A_{g}+20A_{u}+19E_{g}+20E_{u}\]
Thus, there are 38 Raman-active phonon modes
($19A_{g}+19E_{g}$) and 40 infrared-active modes
($20A_{u}+20E_{u}$). It is clear from this analysis that one
should expect a dramatic increase in the number of modes in the low
temperature phase. Some of these new modes are the former
Brillouin-zone-boundary modes which become visible because of the increase of
the unit cell.

\section{Experimental}

The  \textit{1T}-TaS$_2$  single crystals used in this work were
grown by chemical vapor transport. This procedure yielded  single
crystals with a typical size of 10$\times$10$\times$0.2~mm$^{3}$.
Optical measurements were carried out on the optically smooth
surfaces of the as-grown single crystals.

The reflectance was measured in the frequency range of
30--45,000~cm$^{-1}$ (6~meV--5.5~eV) at several temperatures
between 30 and 360~K. The sample temperature was maintained by
mounting the crystal on the cold-finger of a He-flow cryostat. To
cover a wide frequency range, we used three optical setups. A
Bruker IFS 113V spectrometer was used for the far-infrared and
midinfrared regions (50--4,500~cm$^{-1}$; 6--550~meV), a
Perkin-Elmer 16U grating spectrometer were used for the near
infrared and visible region (3,800--20,000~cm$^{-1}$; 0.5--2.5
eV). We observed no temperature dependence of the spectra above
20,000~cm$^{-1}$. Therefore, in order to extend the data to higher
frequencies (20,000-45,000~cm$^{-1}$; 2.5--5.5~eV) we merged the
spectrum at a given temperature with the room temperature
reflectance spectrum obtained from a Zeiss grating spectrometer
coupled with a microscope.

In order to analyze the optical properties of the sample we
performed Kramers-Kronig transformation of the reflectance data.
For  temperatures above 360~K, we used the  Drude-Lorentz model
fit in the low frequency part of the spectrum. We calculated the
reflectance and used the result for the low-frequency
extrapolation. Below the transition, we assumed constant DC
conductivity at zero frequency as a low frequency approximation.
For the high frequency approximation, we used a week power law
followed by free-electron-like behavior.

\section{RESULTS AND DISCUSSION}

\subsection{Low frequency optical conductivity
and infrared-active phonons}

\begin{figure}[h]
\includegraphics{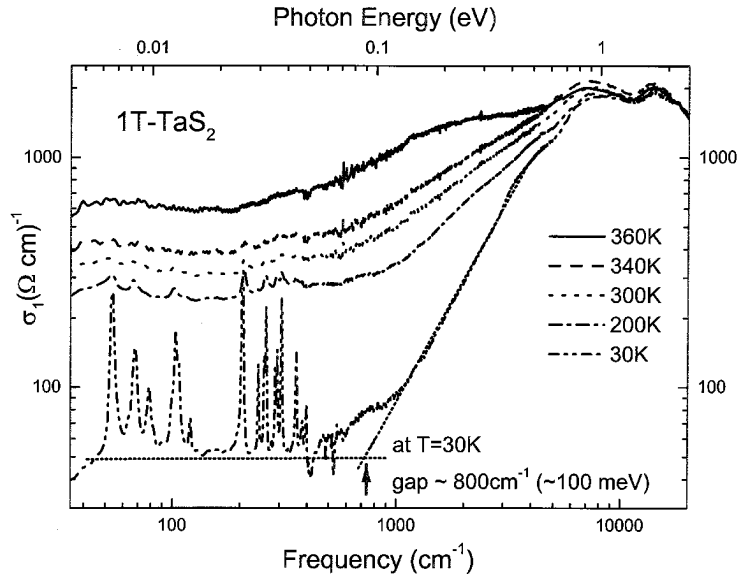}
\caption{\label{fig2} Temperature dependence of the optical
conductivity of \textit{1T}-TaS$_2$. The gap at 800~cm$^{-1}$ is
defined as the onset of the increase of the optical conductivity
in 30K spectrum. The value of the gap is estimated from the
location of the intercept of the background low frequency optical
conductivity  and the line representing increase of the optical
conductivity (dashed lines in the 30~K spectrum).}
\end{figure}

The optical conductivity of  \textit{1T}-TaS$_2$ is  shown in
Fig.\ref{fig2} and \ref{fig3}. Notice the strong changes in both
the low-frequency optical conductivity as well as in the phonon
spectrum. The behavior of the low frequency conductivity is
consistent with a decrease of dc conductivity due to the MI
transitions, similar to earlier reported data\cite{Uch81}.  At
360~K the phonon modes are strongly screened by free carriers.
Only modest traces of the phonon lines are noticeable. The phonon
modes are followed by a pseudogap-like increase of the optical
conductivity, with direct optical transitions observed at
frequencies above 1eV. Observed changes of charge dynamics in
\textit{1T}-$TaS_2$ will be discussed in forthcoming publication.

A decrease of the temperature below 360~K leads to a decrease of
the low-frequency optical conductivity. A similar decrease is
observed in dc-conductivity.\cite{Pillo01}  With this decreased
free-electron Drude-conductivity  the phonon modes become more
pronounced. At 200~K one can clearly see at least eleven phonon
modes at 54, 67, 102,113, 208, 258, 290, 305, 354, 376, 390. This
number is much bigger than the expected two modes in the
near-commensurate CDW phase, which could be related to local
C$_{3i}^1$ symmetry already established in this phase.

As the temperature is lowered below $\approx$180~K, the low
frequency conductivity drops to about one fifth of its value at
room temperature; phonon peaks become clearly resolved and the
pseudogap develops into a gap. We define this gap as the onset of
the increase of the optical conductivity. In Fig.\ref{fig2}, we
estimate the value of the gap from the location of the intercept
of the background low frequency optical conductivity at 30~K
(50~$\Omega^{-1}cm^{-1}$) and line representing increase of the
optical conductivity (liner dependence on the log/log plot of
Fig.\ref{fig2}). From our data we estimate the gap to be of the
order of 800~cm$^{-1}$(100~meV).

At 30~K we clearly see at least 17 phonon modes, at 54, 67, 78,
99, 103, 106, 120, 205, 241, 256, 261 , 287, 292, 306, 356, 379
and 395~cm$^{-1}$. At higher temperatures some of the modes become
difficult to separate. For instance only in the low temperature
spectra can we separate the triplet at  99, 103, and
106~cm$^{-1}$, the doublet at 256 and 261~cm$^{-1}$, and the
doublet at  287 and 292~cm$^{-1}$. The two modes at 54 and
208~cm$^{-1}$ have the highest intensity. Note that the number of
the modes observed in commensurate CDW phase is about half of what
was expected from symmetry analysis.

The phonon modes are visibly separated into two groups,
Fig.\ref{fig2}. The first group occurs at frequencies below
130~cm$^{-1}$ and the second group at the frequencies above
190~cm$^{-1}$. Because Ta atoms are about five times heavier than
S atoms, we assign the low frequency group to Ta vibrations and
the upper to S vibrations.

We fitted all the observed phonon modes with Lorentzian lineshapes
in order to obtain their frequency and linewidth. Than we plotted
the frequency of the modes versus temperature and fitted these
graphs with straight lines. Such an analysis allowed us to
estimate the  magnitude of the phonon frequency change with the
temperature. We considered a change of 2~cm$^{-1}$ over 300~K
($\approx$ 7x10$^{-3}$~cm$^{-1}$/K) as the threshold for our
analysis. Therefore, only the slopes higher than this threshold
are listed in Table \ref{tab:slopes}.

\begin{table}
\caption{Slopes of the temperature
dependance of the phonon mode frequencies in \textit{1T}-TaS$_2$.}
\label{tab:slopes}
\begin{ruledtabular}
\begin{tabular}{cc}
\footnote{Only phonons with the frequency change larger than
2~cm$^{-1}$ per 300~K are listed}
phonon frequency [cm$^{-1}$]  & slope [cm$^{-1}$/K]\\
\hline
  54  & -0.012\\
  104 & -0.013\\
  120\footnotemark[2] & -0.0158\\
  205 &  0.022\\
  257 & -0.011\\
  261 & -0.010\\
  287\footnotemark[2]& -0.008\\
  356 & -0.013\\
  \end{tabular}
\end{ruledtabular}
\footnotetext[2]{This phonon is observed only in the commensurate
CDW phase}
\end{table}

All the modes other than the 208~cm$^{-1}$ mode display the usual
decrease of frequency with increase of temperature (negative
slope). In contrast, the 208~cm$^{-1}$ mode clearly shows
anomalous behavior. As the temperature decreases the frequency of
this mode decreases, i.e., this mode softens as shown in
Fig.\ref{fig4}. This overall change in frequency with temperature
is one of the strongest among the phonon modes in
\textit{1T}-TaS$_2$; see Table \ref{tab:slopes}.

The strongest change of frequency of the 208~cm$^{-1}$ mode
coincides with the 180~K metal-to-insulator transition,
Fig.\ref{fig4}. Therefore, we suggest that this mode is directly
coupled to structural changes at the transition. Fazekas and
Tosatti\cite{Fazekas79} proposed that the CDW formation affects
not only the Ta layer but entire  S-Ta-S sandwich. It was
suggested that sulphur sheets bulge out at the star center. We
propose that such a distortion should have an effect on the
corresponding phonons. This hypothesis, however, should be
corroborated by theory.

A somewhat similar softening of the Raman-active B$_{1g}$ phonon
mode
\cite{Cooper88,Sugai89,Cardona90,Cardona91,Boekholt91,Reznik93,
Leach93,Chen93,Altendorf93,Kakihana96,Devereaux98} was observed in
YBa$_2$Cu$_3$O$_{7-\delta}$ single crystals and was attributed to
the strong coupling of the superconducting gap to this phonon.
\cite{Rashba88,Cardona90,Devereaux95,Devereaux98,ShermanS95,
ShermanB95} We believe that bulging of the sulphur layers in
\textit{1T}-TaS$_2$ may result in a similar coupling to the gap in
this material.

\subsection{The effect of the hysteresis on the spectrum}

\begin{figure}[h]
\includegraphics{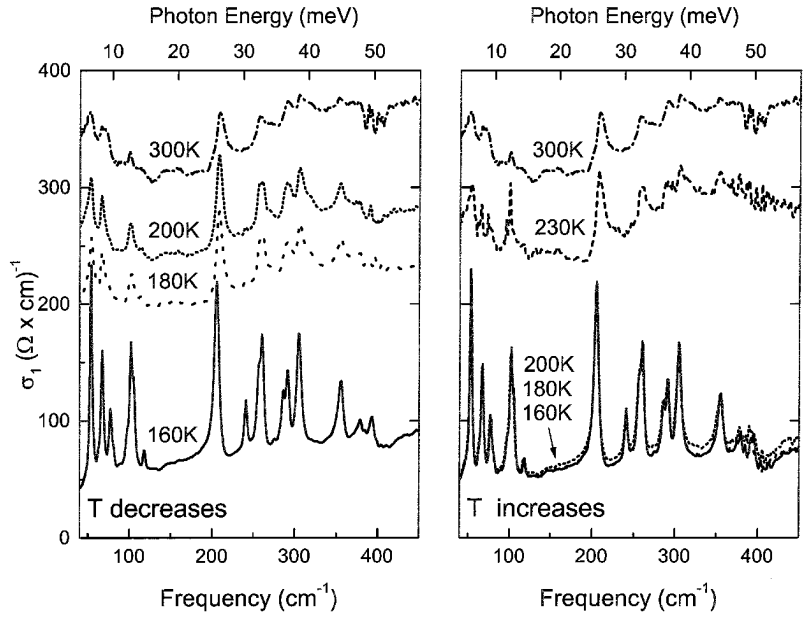}
\caption{\label{fig3} Hysteresis in the optical conductivity of
\textit{1T}-TaS$_2$. The left-hand panel shows the spectra on the
decreasing temperature. The right-hand  panel shows spectra on the
increasing temperature. Note that spectra at 160~K, 180~K, 200~K
on the right-hand panel are practically indistinguishable. The
spectra at the same temperatures but on the decreasing temperature
part of the hysteresis are clearly distinguishable.}
\end{figure}

\begin{figure}
\includegraphics{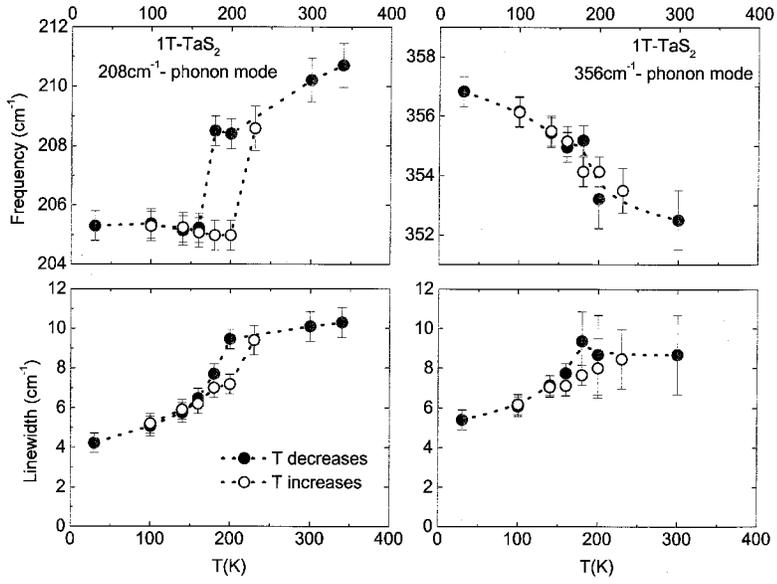}
\caption{\label{fig4} Hysteresis in the frequency and linewidth of
the phonon modes. The left-hand panel shows the frequency (upper
part) and linewidth (lower part) of the 208~cm$^{-1}$ phonon mode
as a function of temperature. The right-hand panel shows the
frequency (upper part) and linewidth (lower part) of the
356~cm$^{-1}$ mode as a function of temperature. The solid circles
correspond to the decrease of the temperature and the open circles
to its increase. The hysteresis is much stronger for the
208~cm$^{-1}$ mode.}
\end{figure}

There is a definite hysteresis in the dc resistivity of
\textit{1T}-TaS$_2$. As a crystal is cooled down, the first MI
transition occurs at 360~K and  the second MI transition occur at
approximately 180~K. However when \textit{1T}-TaS$_2$ is heated
from 30~K the same transition occurs at approximately 220~K.
Figure\ref{fig3} demonstrates the effect of hysteresis on optical
conductivity. The left panel corresponds to decreasing temperature
whereas the right panel represents increasing temperature.

Both the phonon spectrum and the low frequency conductivity are
effected by hysteresis. The effect is especially pronounced for
180~K and 200~K spectra. In particular, the 180~K and 200~K
spectra in the left-hand panel of Fig.\ref{fig3} have
low-frequency conductivities of the order of
250~$\Omega^{-1}$cm$^{-1}$, whereas the corresponding spectra in
the right-hand panel reveal five times smaller values.

The effect of hysteresis is also clear in the temperature
dependence of the frequency and linewidth of the 208~cm$^{-1}$
phonon. The frequency of the mode   on the decreasing-temperature
part of the hysteresis loop is bigger that that measured on the
increasing-temperature part at the same temperature. This
conclusion is also valid for the width  of the mode. We checked
the same effect for all other modes which could be separated above
and below 180~K transition.
 We found that the effect is of the order of
one wavenumber which is within experimental error. Corresponding
data for 356~cm$^{-1}$ mode are shown in the right hand panel of
Fig.\ref{fig4}. To our knowledge this  is the first observation of
such a phonon behavior in \textit{1T}-TaS$_2$.

\section{Conclusions}

We have presented the results of infrared measurements of
\textit{1T}-TaS$_2$. Both the 350~K and 180~K transitions are
accompanied by  strong changes in the low frequency ($\omega <$
1000 cm$^{-1}$) conductivity. We observed  both  additional phonon
modes and an anomalous softening of the 208 cm$^{-1}$ phonon at
the 180~K transition. This effect shows hysteresis similar to that
in dc resistivity. We believe that the softening of the phonon is
directly related to the bulging of sulphur layers that accompanies
the 180~K metal-insulator transition.

\begin{acknowledgments}
Work at the University of North Florida was supported by Research
Corporation Cottrel College Science award CC5290. Work at the
University of Florida was supported by NSF grants DMR-9705108 and
CTS-0082969. The work at the EPFL is  supported by the Swiss F.N.
in part through the  NCCR ``MaNEP". L.V.G would like to thank
E.Ya. Sherman for critical reading of this manuscript.
\end{acknowledgments}
\bibliography{HALK}
\newpage
 \end{document}